\documentclass[twocolumn,showpacs,preprintnumbers,amsmath,amssymb]{revtex4}
\usepackage{tabularx,graphicx}\begin{document}
\newcommand{\beq}{\begin{equation}}
\newcommand{\eeq}{\end{equation}}
\newcommand{\beqn}{\begin{eqnarray}}
\newcommand{\eeqn}{\end{eqnarray}}
\newcommand{\bmath}{\begin{subequations}}
\newcommand{\emath}{\end{subequations}}
\title{Meissner effect, Spin Meissner effect and charge expulsion in superconductors}
\author{J. E. Hirsch }
\address{Department of Physics, University of California, San Diego,
La Jolla, CA 92093-0319}

\begin{abstract} 
 The Meissner effect and the Spin Meissner effect are the spontaneous generation of charge and spin current respectively near the surface of a metal making a transition to the superconducting state. The Meissner effect is well known but, I argue, not explained by the conventional theory, the Spin Meissner effect has yet to be detected. I propose that both effects take place in all superconductors, the first one in the presence of an applied
 magnetostatic field, the second one even in the absence of  applied external fields.  Both effects can be understood under the assumption that electrons expand their orbits and thereby lower their quantum kinetic energy in the transition to superconductivity. Associated with this
 process, the  metal expels negative charge from the interior to the surface and an electric field is generated in the interior. The resulting charge current
 can be understood as arising from the magnetic  Lorentz force on radially outgoing electrons, and the resulting  spin 
 current can be understood as arising  from a spin Hall effect originating in the Rashba-like coupling of the electron magnetic moment to the  internal electric field. The associated electrodynamics  is qualitatively different from London electrodynamics, yet can be described by a small modification of the conventional London equations. 
 The stability of the superconducting state and its macroscopic phase coherence hinge on the fact that the orbital angular momentum of the carriers of the spin current is found to be exactly $\hbar/2$, indicating a topological origin. The simplicity and universality of our theory argue for its validity, and the 
 occurrence  of superconductivity in many classes of materials can be understood within our theory. 
    \end{abstract}
    \date{May 23, 2011}
\pacs{}
\maketitle 

\section{introduction}
The Meissner effect\cite{meissner} is the most fundamental property of superconductors.  I argue that
the Meissner effect is not accounted for by the conventional BCS-Eliashberg-London framework generally believed to explain all aspects
of the superconductivity of conventional  superconductors\cite{tinkham}  (termed `class 1' superconductors in Ref.\cite{cohen}). 
Instead, I   propose that superconductivity  involves fundamental physics that is not described by conventional theory, namely:
(i) superconductors expel negative charge from the interior to the surface\cite{chargeexp}; (ii) it requires dominance of $hole$ carrier transport in the normal state\cite{holesc,holesc2}; (iii)  it is driven by lowering of $kinetic$ $energy$ of the carriers\cite{apparent,apparent2}; (iv) an electric field exists in the interior of superconductors\cite{electrodyn},
(v)
a spin current exists near the surface, in the absence of applied external fields\cite{sm}, and (vi) superconducting
carriers reside in mesoscopic orbits of radius $2\lambda_L$\cite{copses}, with $\lambda_L$ the London 
penetration depth. I argue that the Meissner effect, exhibited by $all$ superconductors, cannot be accounted for  unless  the above listed  effects also exist  in 
superconductors.

\section{the key physical elements}

  \begin{figure}
 \resizebox{8.5cm}{!}{\includegraphics[width=9cm]{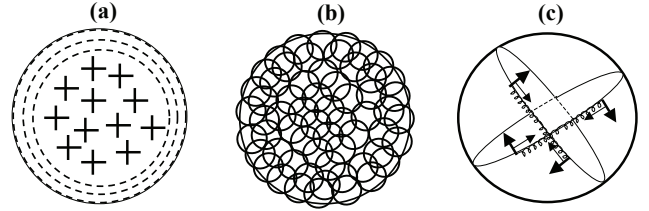}}
 \caption {Illustration of three key aspects of the physics of superconductors proposed here. (a) Superconductors expel negative charge from their
 interior to the region near the surface; (b) Carriers reside in mesoscopic overlapping orbits of radius $2\lambda_L$
 ($\lambda_L$=London penetration depth); (c) A spin current flows near the surface of superconductors (the arrow perpendicular to the orbit denotes the direction
 of the electron magnetic moment). }
 \label{figure2}
 \end{figure}
 
Figure 1 shows three key aspects of the physics of superconductors within the theory discussed here. (a) The charge distribution in the
superconductor is macroscopically inhomogeneous, with excess negative charge near the surface
and excess positive charge in the interior. (b) Superfluid carriers reside in overlapping
mesoscopic orbits of radius $2\lambda_L$. (c) A macroscopic spin current flows near the surface of superconductors in the absence of
applied external fields.

First, why does the Meissner effect imply an inhomogeneous charge distribution as depicted in Fig. 1(a)? Because in order for a
Meissner current to be `spontaneously' generated near the surface of superconductors in the presence of a magnetic field $B$ (in the $z$ direction), 
carriers have to move $radially$ $outward$, to be deflected by the magnetic Lorentz force
\beq
\vec{F}=\frac{e}{c}\vec{v}\times\vec{B}=\frac{e}{c} B(v_r\hat{\theta}-v_\theta\hat{r})\equiv F_\theta \hat{\theta}+F_r\hat{r}
\eeq
in the azimuthal ($\hat{\theta}$)  direction. In the absence of radial motion ($v_r=0$), 
$F_\theta=eBv_r/c=0$, there is no component of the force in the azimuthal
direction and the spontaneous generation of an azimuthal Meissner current cannot be understood.

Second, why does the Meissner effect necessitate orbits of radius $2\lambda_L$? The Larmor diamagnetic susceptibility of $n$ electrons per
unit volume  in 
orbits of radius $r$ for magnetic field perpendicular to the orbits is
\beq
\chi_{Larmor}=-\frac{ne^2}{4m_e c^2} r^2 
\eeq
and takes the value $-1/4\pi$, describing perfect diamagnetism, for 
\beq
r^2=\frac{m_ec^2}{\pi n e^2}
\eeq
from which we conclude that $r=2\lambda_L$, with $\lambda_L$ the London penetration depth, defined by the standard equation\cite{tinkham}
\beq
\frac{1}{\lambda_L^2}=\frac{4\pi n e^2}{m_e c^2}  
\eeq

It is intuitively clear that orbit expansion and charge expulsion are intimately connected: the expanding orbit necessitates that the charge moves
outward. However it is not obvious what is the precise quantitative relation, since it will depend on the degree of overlap of the $2\lambda_L$ orbits.
How is this relation determined?
It turns out that there are  several  very different paths that lead to exactly the same conclusion. This remarkable coincidence strongly suggests
that the result is valid to describe nature.

Let us first state  the results. The electric field in the interior of superconductors is caused by a uniform positive charge distribution. For a long cylinder or a sphere, with radius $R$ much larger than the London penetration depth, it is given by
\beq
E(r)=E_m \frac{r}{R}
\eeq
and is pointing radially outward. Within a London penetration depth of the surface the electric field is screened 
by the excess negative charge near the surface and decays to zero at the surface. 
The maximum electric field, attained for $r \sim R$ is given by\cite{electrospin}
\beq
E_m=-\frac{\hbar c}{4e\lambda_L^2}
\eeq
For $\lambda_L=400A$, $E_m=308,281 V / cm$.
Carriers reside in mesoscopic orbits of radius $2\lambda_L$, and move with speed
\beq
v_\sigma^0=\frac{\hbar}{4m_e\lambda_L}
\eeq
with opposite spin electrons moving in opposite directions. 
Note that 
\beq
v_\sigma^0=-\frac{e}{m_ec}\lambda_L E_m
\eeq
which implies that $v_\sigma^0$ is also the charge velocity that would be generated by a magnetic field of magnitude $E_m$ (in cgs units).
The expelled charge density near the surface ($\rho_{-}$) is related to $E_m$ by
\beq
E_m=-4\pi \lambda_L \rho_{-}
\eeq
so that the expelled charge density screens the interior electric  field. It is also related to the spin current speed $v_\sigma^0$  by
\beq
\rho_{-}=en_s \frac{v_\sigma^0}{c}
\eeq
The orbital angular momentum of the carriers in the $2\lambda_L$ orbits moving at speed $v_\sigma^0$ is 
\beq
L=m_ev_\sigma^0(2\lambda_L)=\frac{\hbar}{2}                 .
\eeq

\section{microscopic derivation}
The spin-orbit interaction derived from Dirac's Hamiltonian for an electron of charge $e$ and mass $m_e$ in an electric
field $\vec{E}$ is
\beq
H_{s.o.}=-\frac{e\hbar}{4m_e^2 c^2} \vec{\sigma}\cdot(\vec{E}\times\vec{p}) .
\eeq
We consider the single-particle Hamiltonian
\bmath
\beq
H=\frac{1}{2m_e}(\vec{p}-\frac{e}{c}\vec{A}_\sigma)^2
\eeq
or equivalently 
\beq
H=\frac{p^2}{2m_e}-\frac{e}{m_e c} \vec{A}_\sigma\cdot\vec{p}+\frac{e^2}{2m_e c^2} A_\sigma^2
\eeq
with the spin-orbit vector potential $\vec{A}_\sigma$ given by\cite{ac}
\beq  
\vec{A}_\sigma=\frac{\hbar}{4m_e c}\vec{\sigma}\times\vec{E}  .
\eeq
\emath
The  term  linear in $A_\sigma$ in Eq. (13b) gives the spin-orbit interaction Eq. (12). In obtaining Eq. (13b) from Eq. (13a), the fact that  
$\vec{\nabla}\cdot(\vec{\sigma}\times\vec{E})=-\vec{\sigma}\cdot(\vec{\nabla}\times\vec{E})=0$ in an
electrostatic situation is used. The significance of the  term quadratic in $A_\sigma$  in Eq. (13b) will be discussed below.

We propose this Hamiltonian to describe the interaction between the charged superfluid and the compensating ionic
background charge. If $n_s$ is the density of superfluid carriers of charge $e$, the compensating ionic background
has charge density
\beq
\rho_i=-en_s
\eeq
The electric field generated by the charge density Eq. (14) in a cylindrical geometry at distance $r$ from the axis is (using Eq. (4))
\beq
\vec{E}=-2\pi en_s\vec{r}=-\frac{m_e c^2}{2e\lambda_L^2}\vec{r}
\eeq
The spin-orbit vector potential is then
\beq
\vec{A}_\sigma=-\frac{\hbar c}{8e\lambda_L^2}(\vec{\sigma}\times\vec{r})=E_m\frac{\vec{\sigma}\times\vec{r}}{2}  .
\eeq
with $E_m$ given by Eq. (6). 
Thus, the superfluid carriers move in a uniform effective magnetic field   $\vec{B}_\sigma$ given by
 \bmath
 \beq
 \vec{A}_\sigma=\frac{\vec{B}_\sigma\times\vec{r}}{2}
 \eeq
 \beq
\vec{B}_\sigma=E_m\vec{\sigma}
\eeq
\emath
The radius of the cyclotron motion (`magnetic length') associated with the magnetic field $B_\sigma$ in the lowest
Landau level is
\beq
l_{B_\sigma}=(\frac{\hbar c}{|e|B_\sigma})^{1/2}=2\lambda_L
\eeq
which, as discussed in Sect. II, is the radius of the orbits required to give rise to a Meissner effect. We believe this coincidence
is not accidental.

The Hamiltonian term that is quadratic in $A_\sigma$ describes the electrostatic energy cost resulting from the orbit expansion
and associated charge expulsion. From Eq. (13b), (16) and (4) we obtain
\beq
H_{quad}=\frac{e^2}{2m_e c^2} A_\sigma^2=\frac{1}{n_s}\frac{E_m^2}{8\pi} \frac{r^2}{(2\lambda_L)^2}
\eeq
under the assumption $|\vec{\sigma}\times\hat{n}|=1$, which we can write as
\beq
H_{quad}=\frac{1}{n_s}\frac{E_s(r)^2}{8\pi}
\eeq
with
\beq
E_s(r)=\frac{E_m}{2\lambda_L} r .
\eeq
The electric field $E_s(r)$ can be understood as the average electric field resulting from charge expulsion when the orbits
expanded to radius $r$, and the Hamiltonian term $H_{quad}$ is the electrostatic energy density divided by the carrier
density $n_s$, hence the electrostatic energy cost per carrier.

The existence of $2\lambda_L$ orbits can be understood using a semiclassical argument from the fact that they give rise to minimum total energy, assuming that the angular momentum is fixed at value $\hbar/2$ (Eq. (11)), which originates in the topological constraint that the pair wavefunction be single-valued. The Hamiltonian Eq. (13b) is, upon replacing
$\vec{E}$ by the expression Eq. (15)

\beq
H=\frac{p^2}{2m_e}+\frac{\hbar}{2m_e}\frac{r}{(2\lambda_L)^2}(\vec{\sigma} \times\hat{n})\cdot\vec{p}
+\frac{\hbar^2}{8m_e}\frac{r^2}{(2\lambda_L)^4} |\vec{\sigma}\times\hat{n}|^4 .
\eeq
For a circular orbit of radius $r$ and angular momentum $\hbar/2$, $p=\hbar/2r$ and Eq. (22) yields
\beq
H=\frac{\hbar^2}{2m_er^2}+\frac{\hbar^2}{4m_e(2\lambda_L)^2}(\vec{\sigma} \times\hat{n})\cdot\hat{p}
+\frac{\hbar^2}{8m_e}\frac{r^2}{(2\lambda_L)^4} |\vec{\sigma}\times\hat{n}|^4 .
\eeq
Assuming $\vec{\sigma}$ is perpendicular to $\hat{n}$ and minimizing Eq. (23) with respect to $r$ yields
\beq
r=2\lambda_L  .
\eeq
Thus, the fact that the orbits expand from a microscopic radius to radius $2\lambda_L$ can be understood as driven by
lowering of kinetic energy (first term in Eq. (23)) at a cost in potential energy (last term in Eq. (23)) to yield minimal 
total energy. A similar argument explains why the ground state radius of the electron in a Bohr atom is $a_0=\hbar^2/ m_e e^2$.

We assume that the single electron states are governed by the  Hamiltonian Eq. (22) for the value of $r$ giving minimum energy
for the electron orbits, i.e. $r=2\lambda_L$. Hence
\beq
H=\frac{p^2}{2m_e}+\frac{\hbar q_0}{2m_e} (\vec{\sigma} \times\hat{n})\cdot \vec{p} +\frac{\hbar^2 q_0^2}{8m_e}
\eeq
 with $q_0=1/2\lambda_L$. The eigenstates of Eq. (25) are plane waves, with energy dispersion relation
\beq
\epsilon_{k\sigma}=\frac{\hbar^2 k^2}{2m_e}-\frac{\hbar^2}{2m_e} q_0 \vec{k}\cdot (\vec{\sigma}\times\hat{n}) +\frac{\hbar^2 q_0^2}{8m_e} .
\eeq
The speed of carriers of spin $\sigma$ and wavevector $\vec{k}$ is
\beq
\vec{v}_{\vec{k}\vec{\sigma}}=\frac{1}{\hbar}\frac{\partial \epsilon_{k \sigma}}{\partial \vec{k}}=
\frac{\hbar\vec{k}}{m_e} -\frac{\hbar q_0}{2m_e} \vec{\sigma}\times\hat{n}
\eeq
so that the carrier's speed increases or decreases by $v_\sigma^0$ (Eq. (7)) depending on whether
$\vec{\sigma}\times\hat{n}$ is parallel or antiparallel to $\vec{k}$. The dispersion relation Eq. (26) gives
rise to two Rashba bands
\bmath
\beq
\epsilon_k^1=\frac{\hbar^2}{2m_e}(k-\frac{q_0}{2})^2
\eeq
\beq
\epsilon_k^2=\frac{\hbar^2}{2m_e}(k+\frac{q_0}{2})^2  .
\eeq
\emath
The lowest energy band, $\epsilon_k^1$, corresponds to spin orientation parallel to $\hat{k}\times\hat{n}$. These bands are 
shown schematically in Figure 2.

  \begin{figure}
 \resizebox{8.5cm}{!}{\includegraphics[width=9cm]{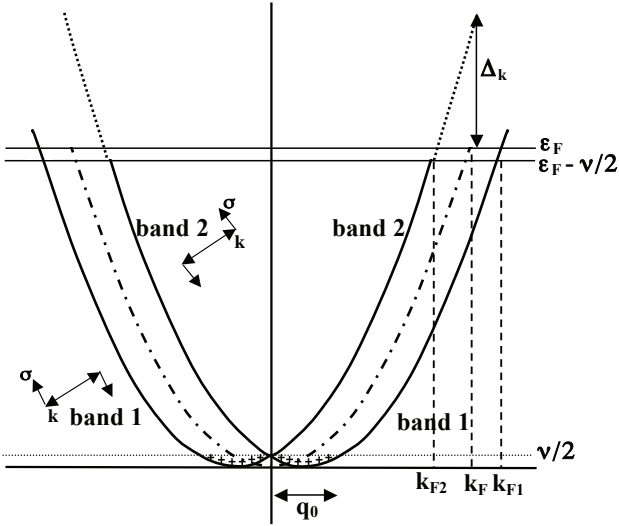}}
 \caption {Rashba bands described by Eq. (28). The Fermi level decreases by $\nu/2=\hbar^2 q_0^2/8m_e$ relative to the
 case where $q_0=0$. The states in band 1 with $k<q_0=1/2\lambda_L$ are unoccupied. The direction of the spin $\sigma$ relative to the
 wavevector is shown schematically for band 1 and band 2, with the convention that the normal to the closest surface points
 $out$ of the paper.}
 \label{figure2}
 \end{figure}
 
 \section{spin current and kinetic energy lowering}

The motion of carriers with spin current velocity $v_\sigma^0$ has associated with it a kinetic energy per carrier
\beq
\epsilon_{kin}=\frac{1}{2}m_e(v_\sigma^0)^2
\eeq
so it may seem that generation of a spin current as the system goes superconducting entails an $increase$ in kinetic energy.
Remarkably, it is exactly the opposite: the system $lowers$ its kinetic energy as it goes superconducting\cite{kinenergydriven}.

One way to see this is as follows: consider a Cooper pair of electrons with opposite spin and spin current
velocity $\pm v_\sigma^0$. When a magnetic field is applied, one of the electrons slows down and the other speeds up by the 
change in  velocity that they acquire, $\Delta v$. The change in kinetic energy of the pair is
\beq
\Delta E=\frac{1}{2}m_e[(v_\sigma^0+\Delta v)^2+(v_\sigma^0-\Delta v)^2]-2(\frac{1}{2} (v_\sigma^0)^2)=m_e (\Delta v)^2
\eeq
so the kinetic energy increases. When one of the components of the spin current stops, i.e. $\Delta v=v_\sigma^0$, the system
goes normal\cite{sm} and at that point the kinetic energy of the pair has increased by
$\Delta E=m_e (v_\sigma^0)^2$, from which we conclude that the
condensation energy per carrier is
\beq
\epsilon_{c}=\frac{1}{2}m_e(v_\sigma^0)^2=\frac{\hbar^2 q_0^2}{8 m_e}
\eeq
or in other words, the carriers $lower$ their kinetic energy by $(1/2)m_e(v_\sigma^0)^2$, rather than raise it by that amount, as they go
superconducting and develop the spin current.

The Rashba bands Eq. (28) describe precisely this physics in a two-dimensional system. The carrier density is given by
\beq
n=\frac{k_F^2}{2\pi}=\frac{k_{F1}^2}{4\pi}+\frac{k_{F2}^2}{4\pi}
\eeq
where $k_F, k_{F1}, k_{F2}$ are the Fermi wavevectors in the state without and with spin current (c.f. Fig. 2).
Since from Eq. (28) $k_{F2}=k_{F1}-q_0$, we have $k_{F1}=\bar{k}_F+q_0/2, k_{F2}=\bar{k}_F-q_0/2$, with
\beq
\bar{k}_F=\sqrt{k_F^2-q_0^2/4}
\eeq
to satisfy Eq. (32). The Fermi energy is lowered from
 \beq
 \epsilon_F=\frac{\hbar^2k_F^2}{2m_e}
 \eeq
 to
 \beq
 \epsilon_{F1,2}=\frac{\hbar^2 \bar{k}_F^2}{2m_e}=\epsilon_F-\frac{\nu}{2}
 \eeq
 with
 \beq
 \nu\equiv\frac{\hbar^2 q_0^2}{4 m_e} .
 \eeq
 
For a constant density of states Eq. (35) implies that each carrier lowers its kinetic energy by $\nu/2$ as the spin current develops in the
superconducting state, hence that the condensation energy per electron is $\nu/2$, in agreement with Eq. (31). This can also be seen directly from the spin orbit interaction Hamiltonian Eq. (12), which
gives rise to the spin orbit interaction energy 
\beq
E_{s.o.}=\frac{e\hbar}{4m_e c^2}\vec{\sigma}\cdot(\vec{v}_\sigma^0\times\vec{E})=\frac{\hbar^2 q_0^2}{4m_e}=\nu
\eeq
for carriers moving with speed $v_\sigma^0$ given by Eq. (7), and the 
 electric field Eq. (15) evaluated at $r=2\lambda_L$ giving  $E=m_e c^2/|e|\lambda_L$. 
 The reason Eq. (37) is twice as large as Eq. (31) is that it does not include the electrostatic energy cost $E_{el}$ 
 \beq
 E_{el}=\frac{1}{n_s} \frac{E_m^2}{8\pi}=\frac{\nu}{2}
 \eeq
arising from the electrostatic field that develops due to the orbit expansion and associated charge expulsion.
In other words, electrons lower their kinetic energy by $\nu$ but give back half of that gain in the electrostatic energy
cost associated with charge expulsion and spin current development.

In fact, the density of states associated with the energy dispersion relations Eq. (28) is not constant as would be the case in an ordinary
two-dimensional system, rather it is given by (per spin per unit area)
\beq
g(\epsilon)=\frac{m_e}{2\pi \hbar^2} \frac{k}{|k\pm\frac{q_0}{2}|}
\eeq
so it is only approximately constant for $k>>q_0/2$. This implies that when the Fermi level drops by $\nu/2$ the energy lowering per particle is not exactly
$\nu/2$. A direct calculation yields for the change in energy in the presence of spin splitting
\beq
\Delta E =N[-\frac{\hbar^2 q_0^2}{8m_e}+\frac{\hbar^2 q_0^4}{48 m_e k_F^2}] .
\eeq
Note that the correction term is very small, a fraction $\sim 10^{-6}$. Nevertheless it has an interesting interpretation. The energy of the electrons in the 
lower Rashba band in the range $k<q_0$ is found to be
\beq
E_{core}=N\frac{\hbar^2 q_0^4}{48 m_e k_F^2}  .
\eeq
We have proposed in earlier work\cite{holecore} that the expelled electrons giving rise to the internal electric field are precisely those in the lower
Rashba band with $k<q_0$, giving rise to a `hole core' of unoccupied states (holes) of long wavelength. 
This is shown schematically in Fig. 2. Since the states are unoccupied 
their energy has to be substracted from that given by Eq. (40), cancelling the second term in Eq. (40) and thus giving rise to energy lowering per electron of precisely $\nu/2$, in agreement with Eq. (31).

The holes occupying the bottom of the lower Rashba bands have a `Fermi surface' at $k=q_0$ describing orbits of real space
radius $1/q_0=2\lambda_L$
and the associated Fermi velocity
is precisely the spin current velocity Eq. (7). 

In summary, we have seen in the last two sections that the orbit expansion, kinetic energy lowering, negative charge expulsion and
spin current development proposed to take place as a metal undergoes a transition to the superconducting state within the theory of hole
superconductivity, can all be understood from the   assumption that the magnetic moments of the electrons in the superfluid interact with the compensating positive charge
of the ions through the spin orbit interaction resulting from Dirac's Hamiltonian.

\section {spin electrodynamics}

The physics described in the previous sections has a simple and consistent description in terms of electrodynamic equations of the same
type as London's equations, without any reference to a microscopic Hamiltonian. These equations were derived\cite{electrodyn,electrospin} before their
microscopic origin was fully elucidated by requiring that the electrodynamic equations be relativistically
covariant, and the fact that their predictions coincide with the physics resulting from the microscopic
Hamiltonian discussed in the previous section strongly suggest that they describe reality.
 These equations describe the flow of a spontaneous spin current within a surface layer of thickness $\lambda_L$, as shown
 schematically in Fig. 3. 

In the charge sector, the electrodynamic equations follow from the assumption that the second London equation
is valid in the form proposed by London
\beq
\vec{J}=-\frac{c}{4\pi \lambda_L^2}\vec{A}
\eeq
with the vector potential $\vec{A}$ obeying the Lorentz rather than the London gauge as in the conventional theory. This is not a statement that violates gauge invariance
but a statement about the physics of the system, which adopts a particularly simple form in the gauges just described. Both descriptions
of the physics can be expressed in gauge invariant form, and it is only experiment that can ultimately decide which one describes
nature.

The electrodynamics in the charge sector is described by the four-dimensional vector equation\cite{electrodyn}
\beq
J-J_0=-\frac{c}{4\pi \lambda_L^2}(A-A_0)
\eeq
with
\bmath
\beq
J=(\vec{J},ic\rho)
\eeq
\beq
A=(\vec{A},i\phi)
\eeq
\emath
where $\rho$ is the charge density, and $\phi$ the electric potential, and 
\bmath
\beq
J_0=(0,ic\rho_0)
\eeq
\beq
A_0=(0,i\phi_0)      .
\eeq
\emath
$\phi_0$ is the electric potential originating in the uniform charge distribution $\rho_0$ that gives rise to the electric field
$E_m$ (Eq. (6)) near the surface. 

  \begin{figure}
 \resizebox{7.5cm}{!}{\includegraphics[width=9cm]{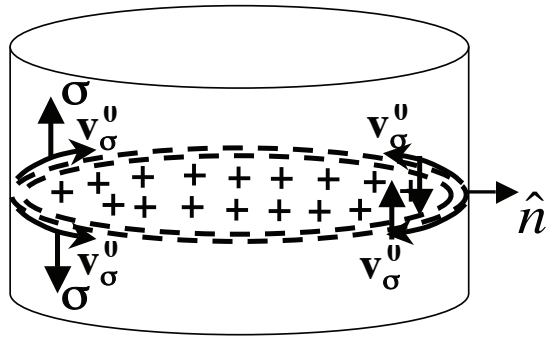}}
 \caption {Schematic depiction of spin current near the surface of a cylindrical superconductor. Superfluid electrons within a 
 London penetration depth of the surface flow counterclockwise (clockwise) if their spin is pointing up (down), with 
 speed $v_\sigma^0$ given by Eq. (7) .}
 \label{figure2}
 \end{figure}

Including spin, the equations are, in a cylindrical geometry with the spin quantization axis parallel to the cylinder axis\cite{electrospin}
\beq
J_\sigma-J_{\sigma 0}=-\frac{c}{8\pi \lambda_L^2}(A_\sigma-A_{\sigma 0})
\eeq
with the four-vectors defined as in Eq. (44), with
\bmath
\beq
\vec{A}_\sigma=\vec{A}+\lambda_L\vec{\sigma}\times\vec{E}
\eeq
\beq
\phi_\sigma=\phi-\lambda_L\vec{\sigma}\cdot\vec{B}
\eeq
\emath
and $J=J_\uparrow+J_\downarrow $, $\rho=\rho_\uparrow+\rho_\downarrow $,  $\phi_{\sigma 0}=\phi_0$    and
\beq
A_{\sigma 0}=\lambda_L\vec{\sigma}\times\vec{E}_0
\eeq
where $\vec{E}_0(\vec{r})$ is the electric field generated by the uniform charge density $\rho_0$ in the interior, that has
magnitude $E_m$ near the surface. Eq. (47b) follows from Eq. (47a) by requiring that the four-divergence of $A_\sigma$ vanishes
and using the Lorentz gauge condition.

Note that the electric part of the vector potential $\vec{A}_\sigma$ is of the same form as Eq. (13c), with the replacement
\beq
r_q\equiv \frac{\hbar}{2m_e c}\rightarrow 2\lambda_L
\eeq
$r_q$ is the 'quantum electron radius'\cite{copses}. 
The transition to superconductivity can be understood as an $expansion$ of the electronic wavefunction from the quantum electron
radius scale $r_q$  to the $2\lambda_L$ scale, keeping the angular momentum fixed at $\hbar/2$, driven by kinetic
energy lowering\cite{kinenergydriven}. The electric field in Eq. (13c) is the bare electric field arising from the charge density
$|e|n_s|$, while the electric field in Eq. (47a) is the net electric field resulting from charge expulsion, which is much smaller.
Their ratio is $r_q/2\lambda_L=v_\sigma^0/c\sim 10^{-6}$.

In the absence of applied magnetic field, Eq. (43) yields for the spin current
\bmath
\beq
\vec{J}_\sigma=-\frac{c}{8\pi \lambda_L}\vec{\sigma}\times(\vec{E}-\vec{E}_0)   .
\eeq
In terms of the superfluid density $n_s$ and the spin current velocity $\vec{v}_\sigma$ we have
\beq
\vec{J}_\sigma=\frac{e n_s}{2}\vec{v}_\sigma
\eeq
\emath
The electric field $\vec{E}$ approaches $\vec{E_0}$ in the interior of the superconductor (at distances larger than $\lambda_L$ from the
surface) and hence the spin current decays to zero in the interior. Near the surface $\vec{E}$ approaches zero as the interior
field $\vec{E}_0$ is screened by the expelled charge density $\rho_-$. At the surface $\vec{E}=0$, $|\vec{E}_0|=E_m$ (Eq. (6))
 and Eq. (49) yields for the
spin current velocity, using Eq. (4)
\beq
\vec{v}_\sigma=-\frac{\hbar}{4 m_e \lambda_L}\vec{\sigma}\times\hat{n}
\eeq
with the normal unit vector $\hat{n}$ pointing outward. This agrees with the spin current velocity 
Eq. (27) derived from the microscopic Hamiltonian Eq. (13).

\section {discussion}

 The theory discussed here offers a new conception of the phenomenon of superconductivity, which naturally ties together 
 many well-known aspects of the physics of superconductors in a very different and more fundamental way than the conventional theory does.
 
   \begin{figure} [b]
 \resizebox{8.5cm}{!}{\includegraphics[width=9cm]{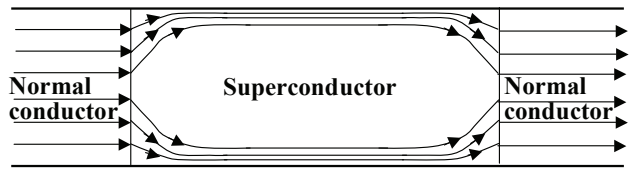}}
 \caption {The current distribution in a superconducting wire which is fed by normal conducting leads.
 The flow lines are calculated in Ref. \cite{londonbook}.
Note that as electrons enter the superconducting region, their velocity acquires a radial component and charge moves towards the surfaces.
In the process they carry with them the magnetic field lines (not shown) which exist throughout the interior in the normal leads
(circles perpendicular to the plane of the page, with normal in the direction of current flow) and only near the surface (and
of course outside the wire) in the superconductor region. }
 \label{figure2}
 \end{figure}

 Superconductivity in our view results from the expansion of the electronic wavefunctions, driven by kinetic energy lowering.
 This tendency of quantum particles to expand 
 their spatial range (generally ascribed to the uncertainty principle) can be easily understood semiclassically: an orbiting particle with fixed angular momentum lowers its
 kinetic energy as the radius of the orbit increases. The need for macroscopic phase coherence follows naturally from the fact
 that the expanded orbits overlap and thus require phase coherence to avoid collisions that would increase the potential energy.
 The process of nucleation of the superconducting state in a normal metal matrix naturally leads to the observed physics within
 our conception: as the orbits expand and the superconducting kernels expands their size, the magnetic field is pushed out of the
 superconducting regions because of the azimuthal currents induced by the Lorentz force acting on the radially outflowing
 charge.  The charge expulsion (which occurs whether or not a magnetic field is present)
  naturally leads to the macroscopically inhomogeneous charge distribution depicted in Fig. (1a), as the 
 expelled charge has nowhere to go but the surface if the entire sample becomes superconducting, in the process carrying the
 expelled magnetic field lines with it. Alternatively, some charge will flow 
 into interior normal regions that trap magnetic field lines, as in the intermediate state of type I superconductors or the
mixed (vortex) state of type II superconductors.

 The fact that negative charge flows out in the transition to superconductivity has not yet been directly verified experimentally.
However I suggest that the phenomenon is clearly illustrated in the current flow through a superconducting wire connected
to normal metal leads shown in Fig. 4. Indeed, as the conduction electrons enter the superconducting region they will 
flow towards the surface as shown by the flow lines in Fig. 4, since current only flows near the surface in 
superconductors. There is only a small leap to the conclusion that charge flows to the surface when a metal
goes superconducting even in the absence of current flow\cite{chargeexp}.

The reader may argue that the $sign$ of the charge moving towards the surface is ambiguous in the situation depicted in Fig. 4, since it depends on the sign of the
charge of the current carriers (whether electrons or holes). Fortunately there is no ambiguity:   experiments on rotating superconductors demonstrate   that 
the current in superconductors is $always$ carried by negative electrons\cite{ehasym, holesc2,rotating,dunne}.

Our  proposal that superconductors expel negative charge from their interior to the surface was motivated by the theory of hole
superconductivity\cite{chargeexp1}, and only later did we realize that it also provides an explanation of the Meissner
effect\cite{giantatom,lorentz,lenz,missing,copses}. It should be pointed out however that priority for this idea 
(unbeknownst to this author when refs. \cite{giantatom, lorentz,lenz,missing,copses} were written) belongs to K.M. Koch\cite{koch}, as
clearly spelled out in ref.\cite{kochjusti}: {\it ``Nimmt man n\"{a}mlich an, dass der \"{U}bergang $N\rightarrow S$ in irgendeiner
Weise mit einer Elektronenbewegung vom Innern des Versuchsk\"{o}rpers nach seiner
Oberflache hin verbunden ist,  - und wir werden sofort sehen, dass zur Verwirklichung einer solchen sogar mehrere
M\"{o}glichkeiten bestehen -, so sieht man ein, dass auf diese Weise ein Abschirmstrom bei konstantem Magnetfeld zustande 
kommen kann.''}

The superconducting state envisioned in our theory   is `dynamic', as it involves motion of electrons in mesoscopic orbits (Fig. 1b) and gives rise
to a macroscopic spin current near the surface (Fig. 1c). Thus it is closer to the state of ``kinetic equilibrium'' envisioned by 
London\cite{london2} than the conventional BCS state. The pre-existent spin currents are readily transformed into charge currents when a magnetic field is
applied, analogous to the way the ``virtual precession'' of the electronic angular momentum in an atom is transformed into real precession when a 
magnetic field is applied\cite{serway}.
This `dynamic' ground state is a truly macroscopic quantum state exhibiting quantum zero point motion at the macroscopic
level, and it bears a qualitative resemblance to early descriptions of the superconducting state
by Bloch, Landau, Frenkel, Smith, Born, Cheng, Heisenberg and Koppe  that envisioned
domains of charge currents pre-existing in the superconductor in the absence of applied
magnetic fields\cite{bloch,landau,frenkel,smith,born,heisenberg,koppe}. Furthermore, Frenkel\cite{frenkel}, Smith\cite{smith} and Slater\cite{slater} pointed out
that electronic orbits of radius of order $\lambda_L$ would naturally explain the Meissner effect (this was also unbeknownst to this author when
we proposed the existence of $2\lambda_L$ orbits\cite{sm}).

In his book `Superfluids'\cite{londonbook}, London coined the term ``Meissner pressure''. By that term he described the tendency of 
the superconductor to  $push$ $out$  the magnetic field lines, against the ``Maxwell pressure'' that tries to keep them inside.  There is no intuitive physical explanation of ``Meissner pressure'' within the conventional London-BCS theory of superconductivity 
 however, other than the  abstract concept that it originates in the difference in the free energy densities of the normal
and superconducting states\cite{londonbook}. Thus, the very descriptive concept of ``Meissner pressure'' articulated by London remained just that,
an appealing physical image with no deeper content.
Instead, for us ``Meissner pressure'' has a concrete physical meaning: it is nothing other than 
the ubiquitous $quantum$ $pressure$\cite{emf}, the tendency of quantum particles to expand their spatial range to lower their
kinetic energy, which has as a $consequence$ the outward motion of  any interior magnetic field lines as well as the outward motion of
negative charge. The `proximity effect', whereby the superconducting state expands into neighboring normal metal regions, is
another vivid manifestation of this physics, as is the prediction that some negative charge `seeps out' of the surface of superconductors\cite{chargeexp,giantatom}.

The tendency of a normal metal to expel negative charge from the interior to the surface and become superconducting will be
largest when electronic energy bands have a lot of electrons (almost full bands, resulting in hole-like carriers)
and when the conducting structures have excess negative charge (conduction through a network of closely spaced
anions). The relevance of these concepts to the understanding of superconductivity in various classes of materials
is discussed in ref. \cite{mm}.

\end{document}